\documentclass[runningheads]{svmult}

\usepackage{makeidx}   
\usepackage{graphicx}  
\usepackage{subeqnar}  
\usepackage{multicol}  
\usepackage{physprbb}  
\makeindex             

\begin{document}
\title*{Abundance gradient in Local Group galaxies using Asymptotic Giant 
Branch stars}
\toctitle{Abundance gradient in Local Group 
\protect\newline galaxies using AGB stars}
%
%
\titlerunning{Abundance in Local Group galaxies}
%
\author{Maria--Rosa Cioni\inst{1}}
\authorrunning{Maria--Rosa Cioni}
%
%
\institute{ESO, Karl--Schwarzschild-Str. 2, D--85748 Garching bei M\"{u}nchen, Germany}

\maketitle              

\begin{abstract}
Simultaneous $IJK_s$ observations allow us to statistically select AGB
stars, both M and C type,  from RGB and younger or foreground stars in
nearby  galaxies.  Regions  of different  metallicity
identified from the  distribution of the C/M ratio show a considerable
variation across the surface of the Magellanic Clouds and NGC 6822.
\end{abstract}

\section{Introduction}
Asymptotic  Giant Branch  (AGB)  stars are  useful  indicators of  the
properties of  a galaxy.  They trace the  intermediate--age population
(between $1$  and several  Gyr), are often  the brightest  and isolated
objects and  can be observed  beyond a Mpc. Furthermore,  because there
are  two kinds  of AGB  stars: oxygen-rich  (O-rich or  M-type) and
carbon-rich (C-rich or C-type), their ratio is a powerful indicator
of metallicity.  AGB  stars are found mostly in  Irregular, Spiral and
Elliptical galaxies in the Local  Group. In particular there are about
$30000$ AGB stars in the Large Magellanic Cloud (LMC), about $8000$ in
the  Small Magellanic  Cloud (SMC)  and  about $3000$  in the  central
region of  NGC 6822.  A handful of  AGB stars  are also found  in some
Spheroidal and  Dwarf Spheroidal galaxies.  It is important,  for the purpose
of this  paper to remember  that there are  more C-rich AGB  stars in
metal poor systems.

Since  1981 Pagel  wrote  that metal  abundance  in external  galaxies
exhibits  radial  gradients.  Searches  for  AGB stars  in  the  solar
neighborhood (\cite{mika}), the Magellanic  Clouds (\cite{blan})  
and Baade's Window  (\cite{glas}) have
shown that the  C/M ratio correlates with metallicity  in the sense: a
higher ratio  corresponds to  a lower metallicity.   Theoretically the
C/M ratio  indicates the  metallicity of a  system because in  a metal
poor environment  the location  of the giant  branch shifts  to warmer
temperatures and less  carbon atoms are needed to  form C stars 
(\cite{scmi}, \cite{ibre}).

Studies  of AGB  stars in  the  Magellanic Clouds  (MCs) have  strongly
improved  thanks to optical and near-infrared (near-IR) large scale  
surveys:  DENIS,  2MASS, MACHO,
EROS,  OGLE, MCPS.  Similar surveys of  other Local
Group  galaxies,  especially  in  the  near-IR wave and
monitoring are taking place.

\section{[Fe/H] Abundance}
\subsection{The Large Magellanic Cloud}
Using   the  DENIS  catalogue   towards  the   MCs  (DCMC) \cite{ccat}
 defined  a  photometric  criteria to  select
stars of  a different type  and age. In the ($I-J$,   $I$)  diagram
  AGB   stars,  irrespectively   of  type,   are  easily
distinguished as a plume of objects above the tip  of the red giant
branch  (TRGB -- \cite{ctip}), from  RGB stars  that are
fainter and  from younger stars or foreground  sources, that
have  much   bluer  colours   (\cite{crat}).   The
distribution  of AGB  stars across  the LMC  describes a
smooth and regular elliptical  structure without clear signs of spiral
features.  The same AGB stars,  because of different molecule that
dominate their  atmosphere, are separated  in O-rich and  C-rich at
$J-K_s=1.4$: C-rich stars are redder than O-rich stars. The
separation  is  a function  of  metallicity.  Our $IJK_s$ selection
criteria includes also O-rich AGB stars of early M spectral sub-type
(M0+) located below the $K_s$-TRGB. In addition most AGB
stars with  a thick  circumstellar envelope ($J-K_s>2.2$)  are excluded 
because they are located below the $I$-TRGB; these can be
O-rich as well as C-rich.

\begin{figure}[t]
\begin{center}
\includegraphics[width=0.75\textwidth]{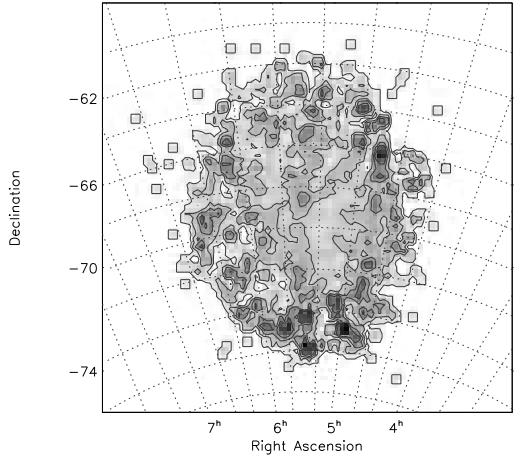}
\end{center}
\caption[]{Distribution of the C/M ratio in the LMC.}
\label{lmcratio}
\end{figure}

The distribution of the C/M ratio (Fig. \ref{lmcratio}) 
is rather clumpy. Regions with an high ratio are progressively located
in the  outer part  of the galaxy  suggesting that the metallicity 
decreases from  the center  to the outer  galaxy. This  gradient has
been recently confirmed by \cite{alve} fitting the RGB in the
($J-K_s$, $K_s$)-2MASS diagram in different locations. In agreement
with \cite{cole}  the bar is  more metal rich  than the
inner  disk region  and  towards  the bridge  connecting  the MCs  the
metallicity is low (Arm B region) as suggested by \cite{stsm}.  
Calibrating  the  C/M  in  terms  of  [Fe/H]  as
discussed in  Sect. \ref{cal}  we obtain a  variation of  $0.75$ dex.
Comparing with the extinction map by \cite{zari} the
C/M0+ ratio is not the result of differential extinction.

\subsection{The Small Magellanic Cloud}
A similar  study, as  presented above,  has been done also for the  SMC 
using DCMC data.  Figure  \ref{smcratio} shows the distribution of the
C/M  ratio.   Here a  clear gradient  is  not present,
however, there are  clumps of high ratio (or  low metallicity) located
in the outer borders of an  inner region, while the outermost parts of
the galaxy have a much lower  ratio. The variation  of {Fe/H] also
corresponds  to $0.75$  dex. An  indication of  a  similar metallicity
distribution has been found  by \cite{zaha}.  The
authors  combined $UBVI$  data from  the Magellanic  Cloud Photometric
Survey   (MCPS)   with    theoretical   isochrones   by \cite{gira}. 
They conclude that s population about  $2.5$ Gyr old 
with $Z=0.008$ is associated to  an outer, perhaps uncertain, ring, the 
latter encloses clumps of objects about $1$--$1.5$ Gyr old with 
$Z=0.001$--$0.004$.

\begin{figure}[t]
\begin{center}
\includegraphics[width=0.75\textwidth]{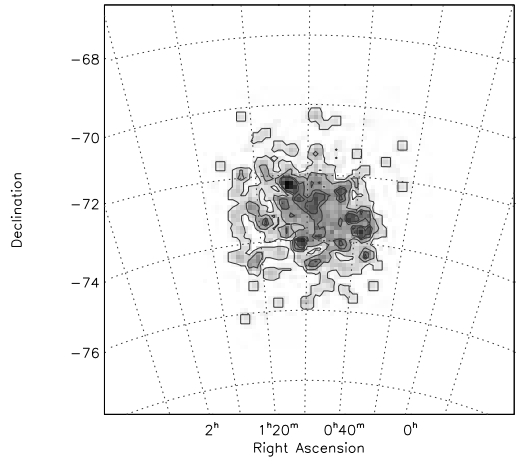}
\end{center}
\caption[]{Distribution of the C/M ratio in the SMC.}
\label{smcratio}
\end{figure}

\subsection{NGC 6822}
NGC6822 is  an isolated  Irregular galaxy in  the Local Group  in many
ways   similar    to   the    MCs   (i.e.   of    intermediate   [O/H]
abundance). Because of  its low latitude it is  affected by a moderate
extinction  ($E(B-V)=0.25--0.45$)   and  contamination  by  foreground
stars. It  started to  form stars about  $10$ Gyr  ago with a  rate that
increased  in the  past $3$  Gyr. Using  $IJK_s$ data  from  the William
Herschel telescope  in La Palma I  and Habing have  surveyed the inner
$20^{\prime}\times 20^{\prime}$  of the galaxy  down to about  $1$ mag
below the TRGB  (\cite{cnew}). These are the
first  near-IR  observations that  cover  the  whole galaxy;  
\cite{davi} observed in $J$ \&  $K_s$ only three very small regions.
Using a similar selection technique as for the MCs we have studied the
distribution   of   AGB   stars   and   of   the   C/M   ratio   (Fig.
\ref{n6822ratio}). The latter corresponds  to a variation of [Fe/H] of
1.89  dex, about  twice as  much as  that found  within the  MCs. This
agrees with  the spread derived  by \cite{tolt} from
RGB stars.  Clumps of high  ratio (or low metallicity) are distributed
in two NS semi-circles around the  central bar, that has on the other
hand a much lower ratio. At least some clumps correspond to regions of
high HI column density (\cite{blwa}).

\begin{figure}[t]
\begin{center}
\includegraphics[width=0.7\textwidth]{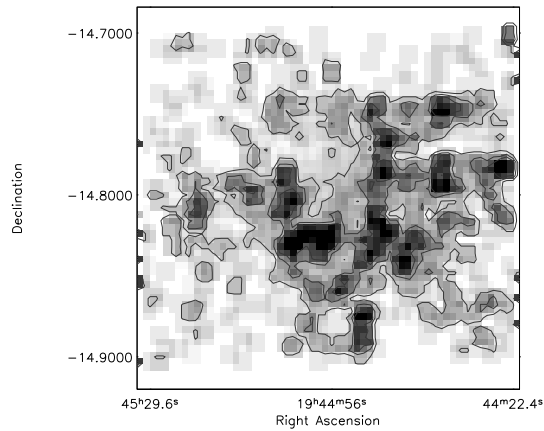}
\end{center}
\caption[]{Distribution of the C/M ratio in NGC 6822.}
\label{n6822ratio}
\end{figure}

\subsection{Calibration of C/M versus [Fe/H]}
\label{cal}
In order to relate the C/M  to [Fe/H] we have used values available in
the    literature     from    the    compilation     by   
\cite{groe} -- see also this proceeding.  
Though a  correlation is  clearly present (Fig. \ref{metal}) 
the  fit is
rather uncertain:  measurements of [Fe/H]  might be in error  by $0.2$
dex while the number of AGB stars  up to 50\%. In fact the former rely
on just a few stars or HII regions and the latter on the extrapolation
to the whole galaxy of the  number of objects detected in small survey
areas often  from incomplete and  inhomogeneous samples.  In  order to
improve this relation we have recently obtained spectra of about $300$
AGB stars in  NGC 6822 to measure the Ca II  triplet and derive, using
the most  up-to-date calibration between these  features and [Fe/H],
an estimate of the metallicity.  These values will be averaged in bins
that contain a  significant number of AGB stars  and hopefully we will
be able to put further constraints on the C/M versus [Fe/H] relation.

It is possible  that the C/M also depends on  the local star formation
history.   In collaboration  with  Girardi, Marigo  \&  Habing we  are
comparing  the  luminosity function  of  AGB stars  in  the  MCs, in 
different locations, with  a  theoretical  luminosity function  to
address this aspect.

\begin{figure}[t]
\begin{center}
\includegraphics[width=0.5\textwidth]{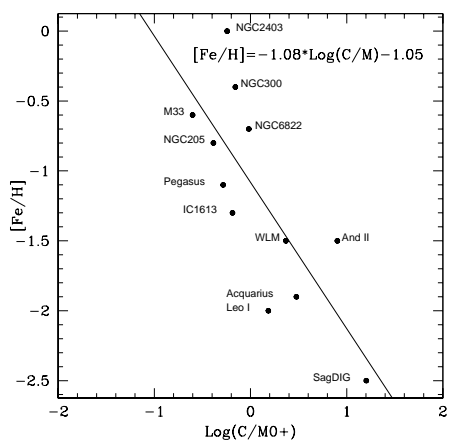}
\end{center}
\caption[]{Relation between the C/M ratio and [Fe/H] in the Local Group.}
\label{metal}
\end{figure}

\section{Other Abundances}
Work in progress on the analysis  of FLAMES spectra of a sample of AGB
stars  in the  LMC will  allow us  to derive  the metallicity  and the
abundance of other  elements that play a key role  in the evolution of
AGB stars.  Preliminary results from data obtained  during the science
verification of the instrument are  very encouraging and were followed 
by new observations in order to  reach a necessary statistical sample
 to study abundances as a function of magnitude, colour and period.

%

\end{document}